\documentclass[aps,prl,superscriptaddress,twocolumn]{revtex4-1}
\usepackage{graphics}
\usepackage{graphicx}
\usepackage{color}
 \usepackage{amsmath, amsthm, amssymb}
 \usepackage{subfigure}
  \usepackage{natbib}

\newcommand{\abs}[1]{\left\vert#1\right\vert}

\begin{document}

\title{Detecting Perfect Transmission in Josephson Junctions on the Surface of Three Dimensional Topological Insulators}

\author{Roni Ilan}
\email{ rilan@berkeley.edu}
\affiliation{Department of Physics, University of California, Berkeley, California 94720, USA}

\author{Jens H. Bardarson}
\affiliation{Department of Physics, University of California, Berkeley, California 94720, USA}

\author{H.-S. Sim}
\affiliation{Department of Physics, Korea Advanced Institute of Science and Technology, Daejeon 305-701, Korea}

\author{Joel E. Moore}
\affiliation{Department of Physics, University of California, Berkeley, California 94720, USA}
\affiliation{Materials Sciences Division, Lawrence Berkeley National Laboratory, Berkeley, CA 94720}


\begin{abstract}

We consider Josephson junctions on surfaces of three dimensional topological insulator nanowires. We find that in the presence of a parallel magnetic field, short junctions on nanowires show signatures of a perfectly transmitted mode capable of supporting Majorana fermions.  Such signatures appear in the current-phase relation in the presence or absence of the fermion parity anomaly, and are most striking when considering the critical current as a function of flux $\Phi$, which exhibits a peak at $\Phi=h/2e$. The peak sharpens in the presence of disorder at low but finite chemical potentials, and can be easily disentangled from weak-antilocalization effects. The peak also survives at small but finite temperatures, and represents a realistic and robust hallmark for perfect transmission and the emergence of Majorana physics inside the wire. \end{abstract}


\maketitle

Three dimensional topological insulators (3DTIs) are materials with a bulk energy gap, supporting surface states with a low energy description of a gapless Dirac fermion, similar to that of graphene~\cite{PhysRevLett.98.106803,PhysRevB.75.121306,PhysRevB.79.195322,3DTIrev1,RevModPhys.83.1057,RevModPhys.82.3045}. Unlike graphene, a 3DTI surface state spectrum may host  an odd number of Dirac cones, making it robust to time reversal symmetry conserving perturbations. While the Dirac spectrum has already been identified in experiments, transport signatures of the non-trivial topological properties of the surface state are more ambiguous~\cite{0034-4885-76-5-056501}. A principal reason for this is the large contribution to transport coming from the bulk, which is not truly insulating~\cite{0034-4885-76-5-056501}. It is therefore highly desirable to identify realistic experimental setups and observables that will provide unique signatures of the surface states, and help characterize them. 

A topological insulator nanowire threaded by magnetic flux $\Phi=\Phi_0/2$, where $\Phi_0=h/e$ is the magnetic flux quantum, is predicted to host a perfectly transmitted mode (PTM)~\cite{JPSJ.71.2753,PhysRevLett.101.086801,PhysRevB.82.041104,PhysRevLett.105.036803}. This mode is a manifestation of the spin-momentum locking typical to TI surfaces and a direct consequence of the cancellation between two phases: a $\pi$ Berry phase accumulated when an electron encircles the wire, and an Aharonov-Bohm phase due to the presence of flux. Transport measurements have already demonstrated the Aharonov-Bohm effect in nanowires threaded by flux~\cite{Peng:2009vw,Xiu:2011wb,Zhao:2011wh,Dufouleur:2012tk}. When the PTM dominates transport (for example, at the Dirac point in a long wire), the wire is expected to show a conductance universally quantized at a value of $e^2/h$~\cite{PhysRevLett.105.156803}. The formation of a PTM in the spectrum occurs since time reversal symmetry is restored at the surface for $\Phi=\Phi_0/2$, but with an odd number of modes (at zero flux the number of modes is even and there is no PTM). As long as time reversal symmetry remains intact, that mode is guaranteed to exist~\cite{0034-4885-76-5-056501}. 

Transport in the presence of a superconductor (SC) offers a good way to disentangle the contributions coming from the bulk and the surface, as well as contributions coming from modes within the surface. Suppression of bulk transport has been observed experimentally~\cite{VenDerWiel}. Once the contribution from the bulk is suppressed, Josephson junctions (JJs) can be used to form a discrete set of Andreev bound states on the surface, separated by an energy scale determined by the SC gap $\Delta$. A JJ with a superconducting phase difference of $\phi=\pi$ situated on the surface of a 3DTI is predicted to host a zero energy Majorana bound state~\cite{PhysRevLett.100.096407,0034-4885-75-7-076501,BeenakkerReview1}. Several recent theoretical works target effects associated with this Majorana bound state~\cite{Grosfeld19072011,FuPotter,arXiv:1302.2113}. This zero energy mode is intimately related to the PTM of the normal region: it is only when the normal region has such a mode that the JJ is able to host a zero energy bound state.  If the normal region between the two superconductors is further gapped (for example, by placing it adjacent to a ferromagnet or terminating the wire altogether), the Majorana modes remain confined at the edge of the SC region~\cite{PhysRevB.84.201105,PhysRevB.86.155431}. Hence JJs on 3DTI nanowires also constitute a promising pathway for observing effects associated with Majorana fermions. 

In this paper, we propose using JJs to explore a surface state behavior unique to 3DTIs. We single out effects associated with the emergence of the PTM and, as a result, zero energy Majorana bound states. The current phase relation (CPR) in the presence of parallel flux through the wire can be a $2\pi$ or a $4\pi$ periodic function, and we show that the $2\pi$ periodic structure displays a characterizing discontinuity at $\Phi=\Phi_0/2$. Furthermore, the critical current as a function of flux has a peak at $\Phi=\Phi_0/2$ with a quantized value at zero temperature. Both features are shown to exist at a finite chemical potential and in the presence of disorder. In particular, the peak shows a remarkable behavior: its presence is \textit{unaffected} by strong non-magnetic disorder, and its features are \textit{sharpened} when the disorder strength is increased. The observation of such a peak in the critical current would provide clear evidence for the formation of the PTM in the junction and a Majorana bound state at a superconducting phase difference of $\pi$. 

To clarify the experimental value of the proposed system, we point out that the main mechanisms that hinder the identification of effects related to Majorana fermions in other systems do not affect the signatures discussed here. Prominent examples of other proposed signatures of Majoranas include the fractional Josephson effect ~\cite{1063-7869-44-10S-S29,FractionalJE,PhysRevLett.100.096407,PhysRevLett.105.177002,PhysRevLett.106.077003,PhysRevLett.103.107002,PhysRevB.79.161408,PhysRevLett.105.077001} and the zero-bias anomaly of SC-normal interfaces in helical wires. The fractional Josephson effect is sensitive to quasi-particle poisoning interfering with the system's ability to maintain a fixed fermion parity. The zero-bias anomaly, recently observed in one-dimensional conventional semiconductor nanowires~\cite{Mourik25052012,Heiblum}, could originate from non-topological sources such as disorder and geometrical effects~\cite{PhysRevLett.109.227005,PhysRevLett.109.267002,PhysRevB.86.100503,1367-2630-14-12-125011}.  In contrast, we show that signatures of the non-chiral (extended) Majorana bound state exist in the CPR (in different forms) in our system with and without parity conservation, while the appearance of a peak in the critical current is indifferent to quasi-particle poisoning. Furthermore, in the present setup disorder strengthens rather than obscures the desired signature of the PTM.

  \begin{figure}[t!]
\begin{center}
\includegraphics[height=1.2in]{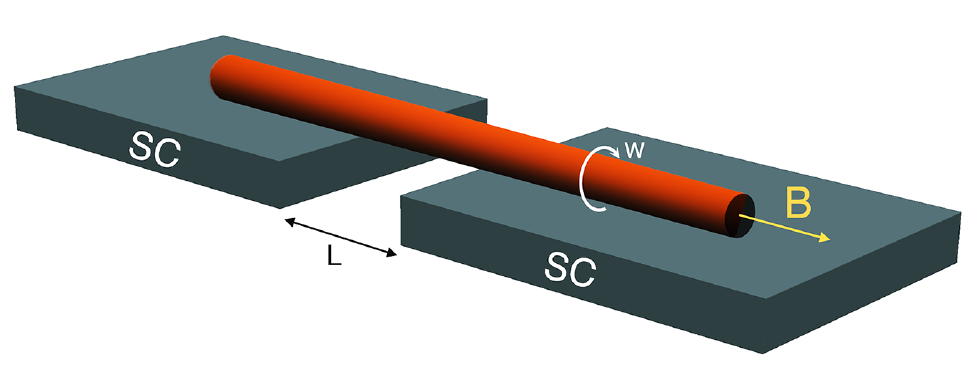}
\end{center}
\caption{Josephson junction on a 3DTI nanowire. The wire has a circumference $W$, and the length of the Josephson junction is defined by the separation between the superconducting leads $L$. We assume $L\ll \xi$, the superconducting coherence length, and $W\lesssim L$.}
\label{fig:setup}
\end{figure}

 We begin by introducing the setup and reviewing the main theoretical components. Consider a 3DTI nanowire placed on top of two superconducting leads as shown in Fig.~\ref{fig:setup}. Define $y$ to be the coordinate that runs along the circumference of the wire $0<y<W$ and $x$ to be the coordinate along the wire. For a clean wire, states can be labeled according to their transverse momenta $q_n$. The quantized values of $q_n$ are determined by the boundary conditions of the Dirac fermion wave functions. In the absence of a magnetic flux through the wire $q_n=2\pi(n+1/2)/W$ with $n=0,\pm1...$, while in the presence of flux, the boundary conditions are shifted due to the Aharonov-Bohm effect, $q_n=2\pi(n+1/2+\Phi/\Phi_0)/W$~\cite{PhysRevLett.105.156803}. When $\Phi=\Phi_0/2$, the set of momentum values contains a state with $q=0$, and therefore the total number of modes is odd. The SC leads induce a superconducting gap $\Delta$ on the surface of the wire for $x<0$ and  $x>L$, forming a JJ of length $L$.  In this paper we consider only short junctions, for which $L\ll \xi$, where $\xi$ is the SC coherence length. In principle, the long junction limit can be studied by generalizing the results of Ref.~\onlinecite{PhysRevLett.110.017003}.

 The transport properties of the wire can be accounted for within a scattering matrix formalism, reviewed in Ref.~\onlinecite{BeenakkerReview2,*Fukuyama:1992uo}
 and applied in Ref.~\onlinecite{PhysRevB.74.041401} for studying JJs on graphene. The critical current through a single graphene sheet is obtained by solving the Dirac - Bogoliubov - de Gennes equation for the SC - normal surface state - SC junction. The energies of the Andreev bound states inside the junction are 

\begin{equation}E_n=\Delta\sqrt{1-\tau_n\sin^2\phi/2}\end{equation}
where $\tau_n$  is the normal state transmission probability for the $n$-th mode
\begin{eqnarray}\label{eq:transmission}
&&\tau_n=\frac{k_n^2}{k_n^2\cos^2(k_nL)+\mu/\hbar v\sin^2(k_nL)}.
\end{eqnarray}
Here $k_n=\sqrt{(\mu/\hbar v)^2-q_n^2}$, $\mu$ is the chemical potential and $v$ the fermi velocity.
The Joesphson current is given by 
\begin{equation}\label{eq:current}
I(\phi)=\frac{e}{\hbar}\sum_n\frac{\partial E_n(\phi)}{\partial \phi}=\frac{e\Delta}{4\hbar}\sum_n\frac{\tau_n\sin\phi}{\sqrt{1-\tau_n\sin^2(\phi/2)}}.
\end{equation}

  \begin{figure*}[t!]
\includegraphics[height=0.18\textwidth]{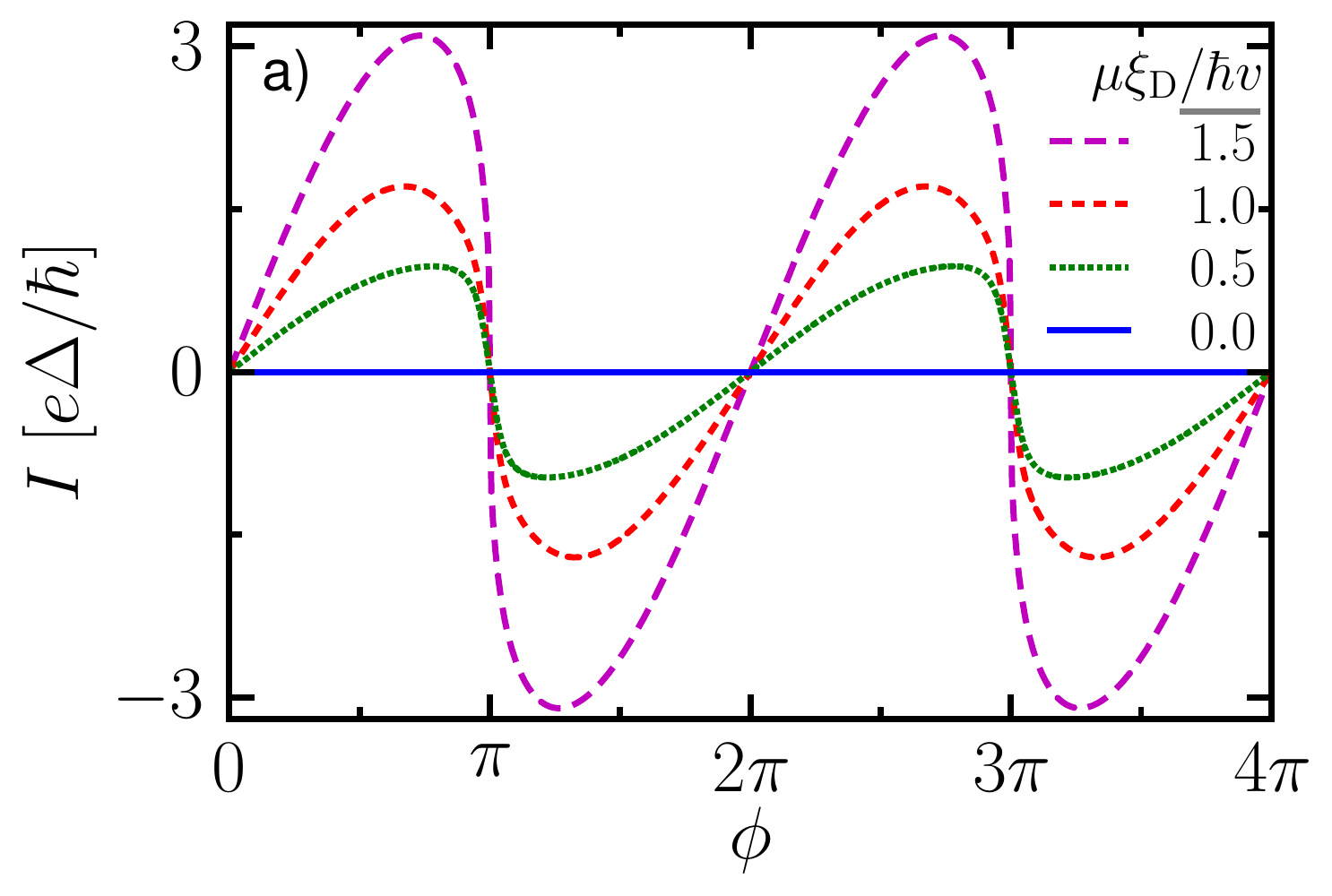}\hspace{-0.2cm}
\includegraphics[height=0.18\textwidth]{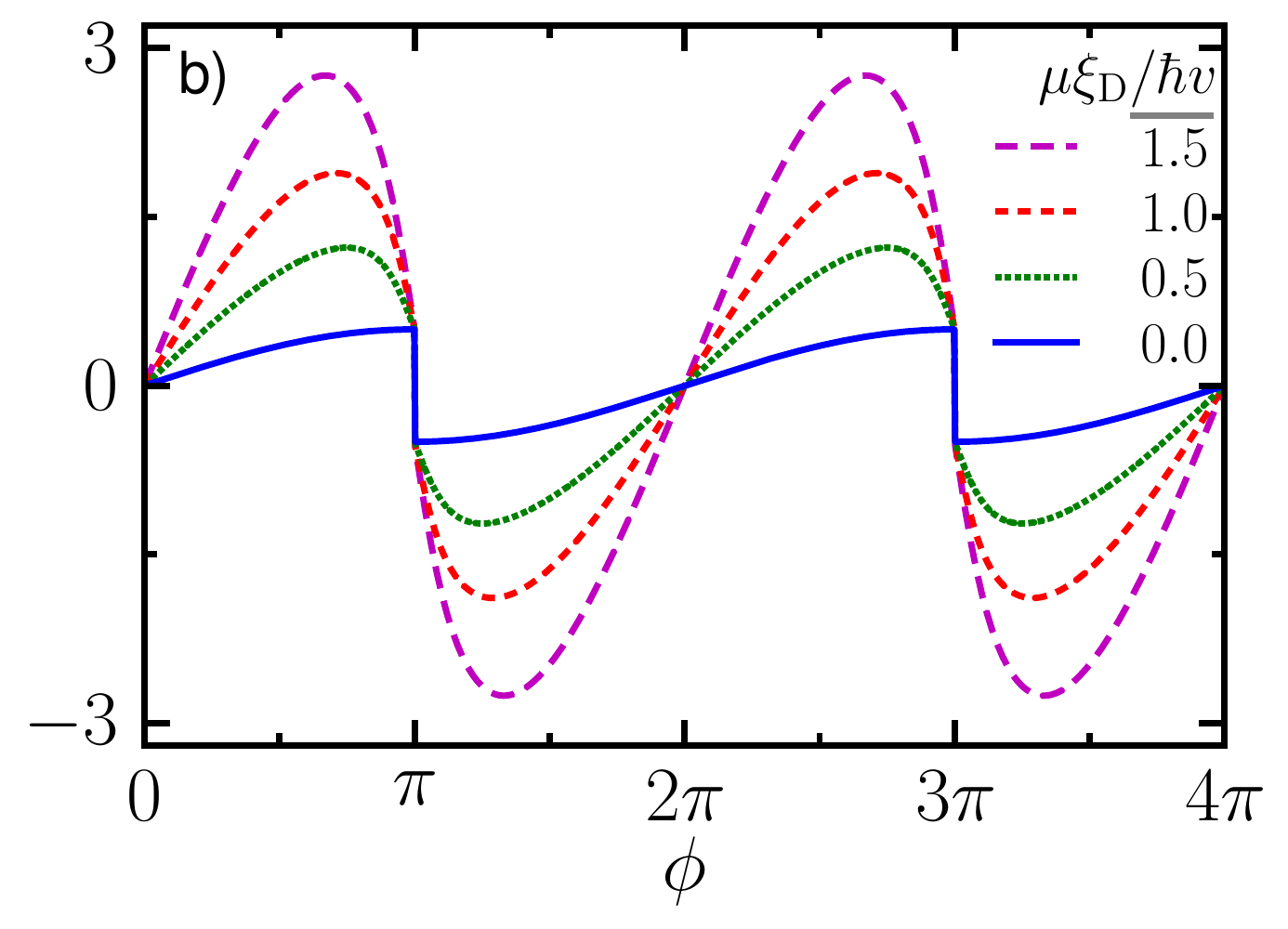}\hspace{-0.2cm}
\includegraphics[height=0.18\textwidth]{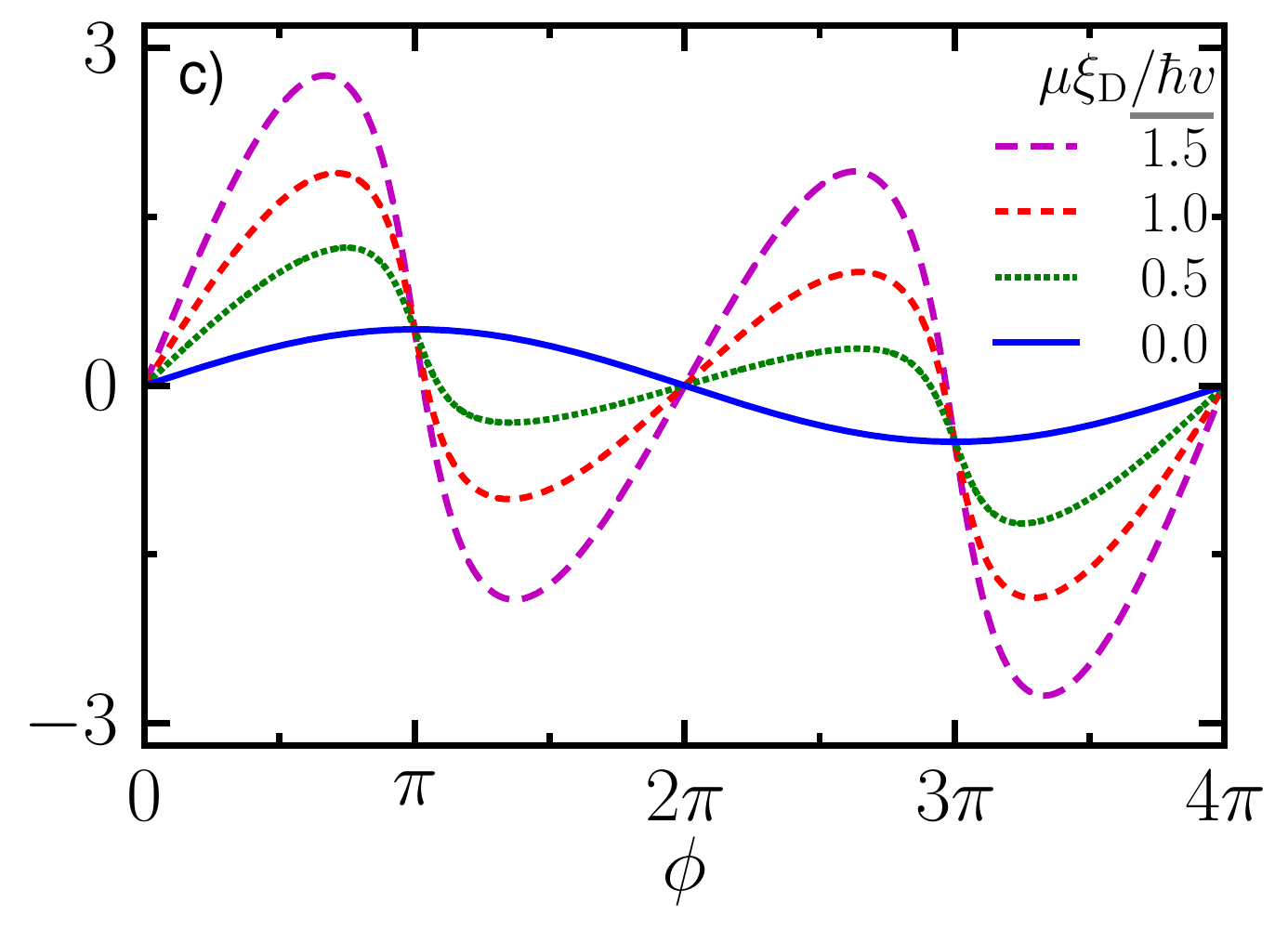}\hspace{-0.2cm}
\includegraphics[height=0.18\textwidth]{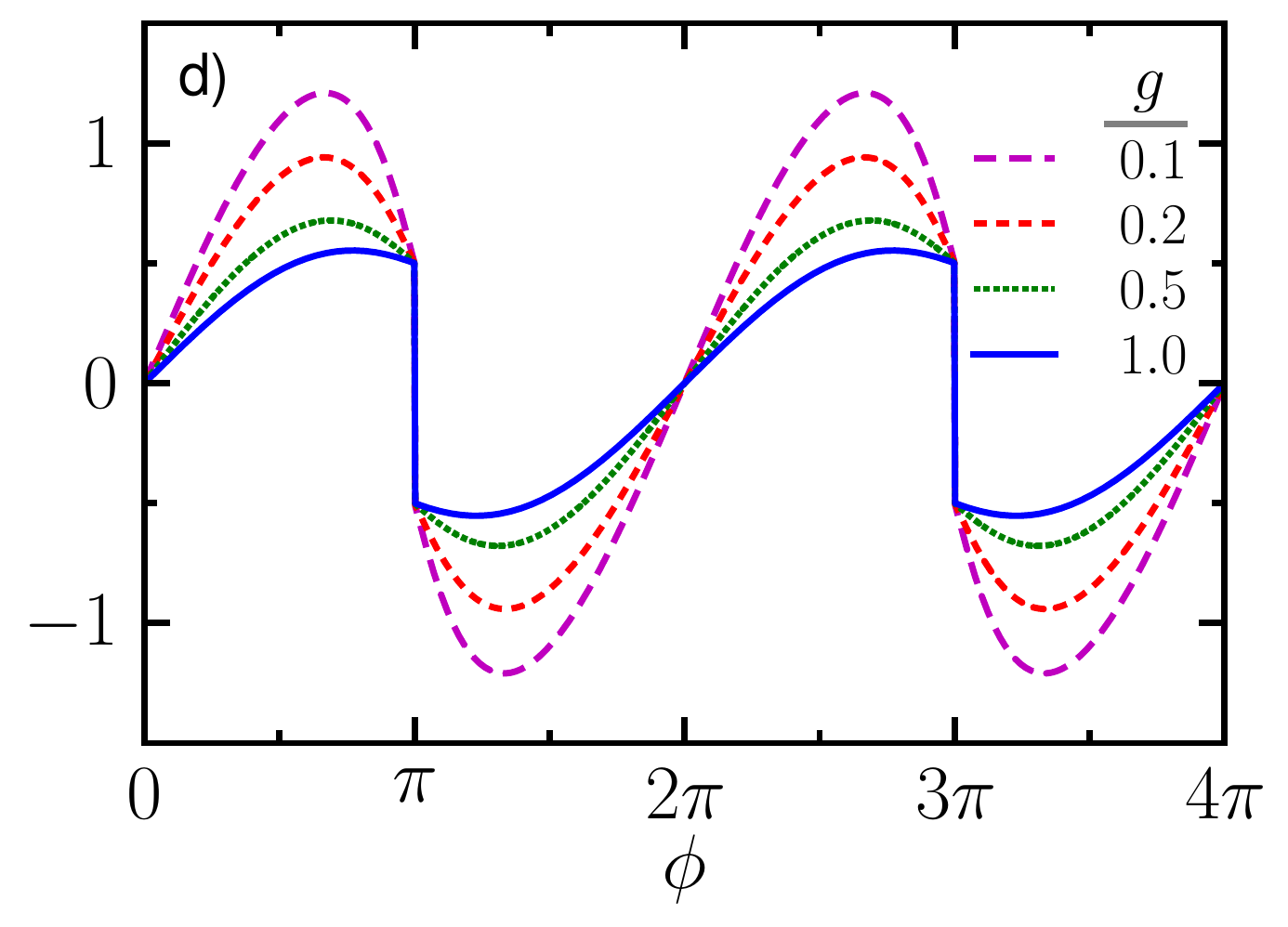}
\caption{Current phase relation. a) CPR for a clean wire at $\Phi=0$ with no parity conservation. b) CPR for a clean wire at $\Phi=\Phi_0/2$ with no parity conservation. c) CPR for a clean with at $\Phi=\Phi_0/2$ in the presence of the parity anomaly. d) CPR for a disordered wire at $\mu\xi_\text{D}/\hbar v=1$, with no parity conservation. For all curves $L/W=4$ ($L/\xi_\text{D}=80 $ and $W/\xi_\text{D}=20 $).}
\label{fig:currentPhase}
\end{figure*}

The transmission probabilities $\tau_n$ are exponentially suppressed in $L/W$ for the modes $q_n$ that render $k_n$ imaginary. Therefore, tuning the chemical potential (or the ratio $L/W$) allows selecting a small number of modes that will contribute to transport.  The possibility of tuning the boundary conditions with the flux now proves to be crucial; when the flux amounts to half a flux quantum, the zero momentum state has perfect transmission, $\tau=1$. The energy for that particular mode remains a function of $\phi$
\begin{equation}\label{eq:majorana branch}
E(\phi)=\pm\Delta\abs{\cos\phi/2},
\end{equation}
and is zero at $\phi=\pi$~\cite{PhysRevLett.100.096407}. 

The current Eq.~\eqref{eq:current} still holds in the presence of disorder, however, the transmission coefficients $\tau_n$ are obtained numerically following Ref.~\onlinecite{PhysRevLett.99.106801} with a Gaussian correlated potential
\begin{equation}
\langle V(r)V(r')\rangle= g \frac{(\hbar v)^2}{2\pi \xi_\text{D}^2}e^{-|r-r'|^2/2\xi_\text{D}^2}.
\end{equation}
Here, $\xi_\text{D}$ is the disorder correlation length, and $g$ a dimensionless measure of the disorder strength. The transmission probabilities are all modified by disorder, apart from that of the PTM, as time reversal symmetry binds its value to unity. Our data is obtained by averaging over $5\cdot 10^{2}-10^{4}$ disorder configurations.

The PTM has a strong signature in transport.  Let us start by considering the CPR $I(\phi)$. Plots of the CPR are shown in Fig.~\ref{fig:currentPhase} for $\Phi=0$ and  $\Phi=\Phi_0/2$ at a fixed value of $L/W=4$ for various values of $\mu$. For $\Phi=0$ the CPR is a $2\pi$ periodic function, with an exponentially vanishing amplitude at $\mu=0 $. As $\mu$ is increased from zero, more modes acquire a finite transmission probability and the CPR oscillates with an amplitude that grows with the number of modes. In contrast, when $\Phi=\Phi_0/2$, a finite amplitude is expected for any value of $\mu$ due to the presence of a PTM. The two sets of curves displaying the CPR at $\Phi=\Phi_0/2$, Fig.~\ref{fig:currentPhase}b and Fig.~\ref{fig:currentPhase}c, are distinguished by the ``fermion parity anomaly'' caused by the change in fermion parity of the ground state at $\phi=(2m+1)\pi$, where $m$ is an integer~\cite{0034-4885-75-7-076501,BeenakkerReview1}. If the system follows the ground state regardless of its parity (Fig.~\ref{fig:currentPhase}b), the current is expected to have a period of $2\pi$, and the particular shape of the low energy band dictated by Eq.~\eqref{eq:majorana branch} forces the current to have a sharp discontinuity at $\phi=(2m+1)\pi$. If the fermion parity cannot change although the ground state parity is changing (Fig.~\ref{fig:currentPhase}c), the current contributed by the PTM has a period  of $4\pi$. In both cases, the contribution of higher energy modes has a period of $2\pi$, and the total current will have a period of $2\pi$ at large $\mu$. Note that the system need not be at the Dirac point to have a periodicity of $4\pi$ in the case where parity is conserved, or a sharp discontinuity in the case where there is no parity conservation.  Both features extend to finite values of $\mu$. 

Given the importance of disorder in current experiments, we now test the robustness of these features.  Disorder tends to localize modes that are not  protected by time reversal symmetry and diminish their contribution to the supercurrent. Hence the effect of increasing the disorder strength at a fixed and finite value of $\mu$ is to reduce the value of the unprotected transmission probabilities. Fig.~\ref{fig:currentPhase}d presents the CPR at $\Phi=\Phi_0/2$ for various values of the disorder strength $g$. Indeed, the amplitude of the current is reduced with disorder until it becomes identical to that contributed by the PTM, and crucially the current discontinuity survives to finite $g$.

\begin{figure}[t!]
	\begin{center}
\includegraphics[width = 1\columnwidth]{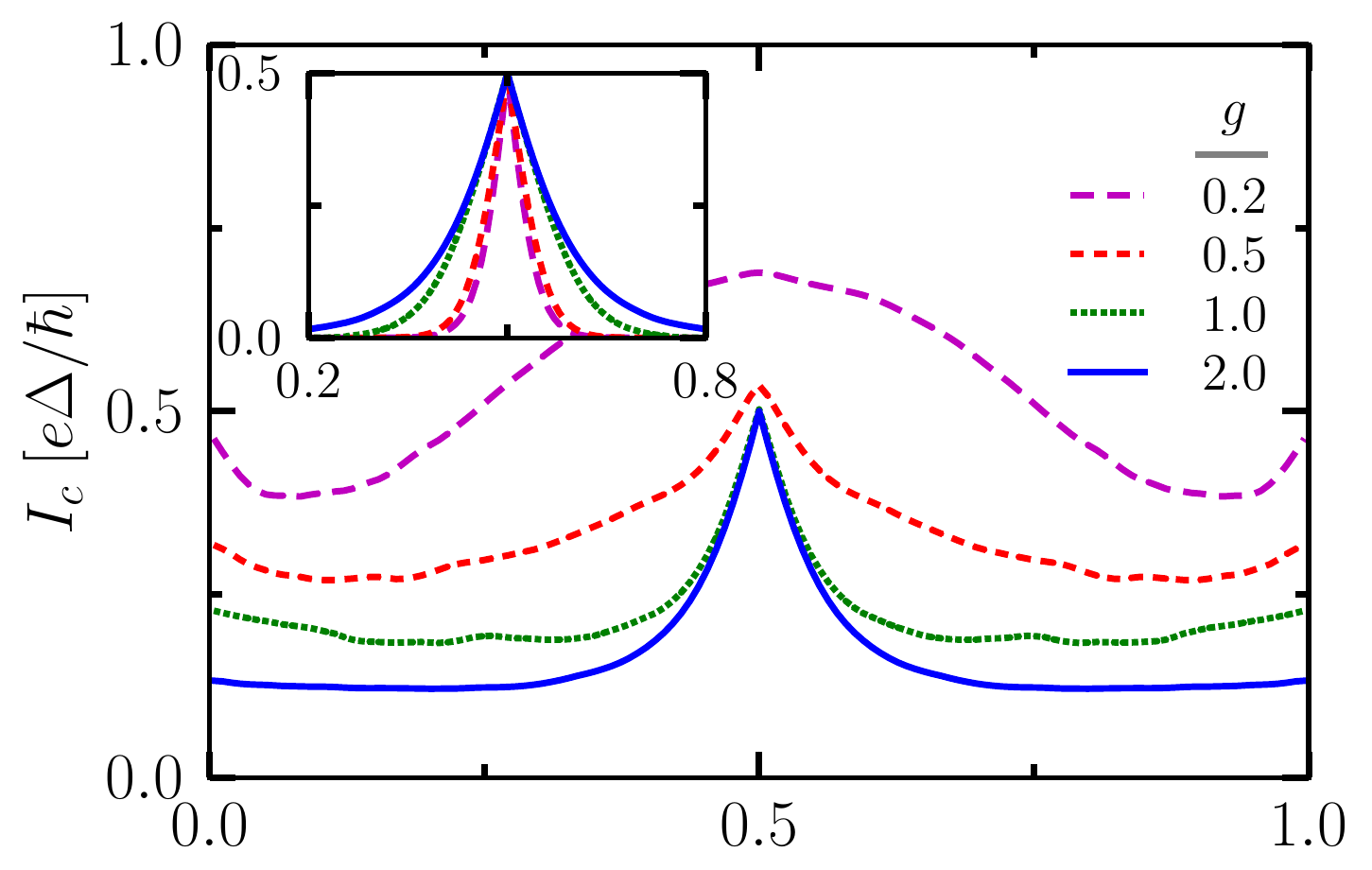}
\includegraphics[width = 1\columnwidth]{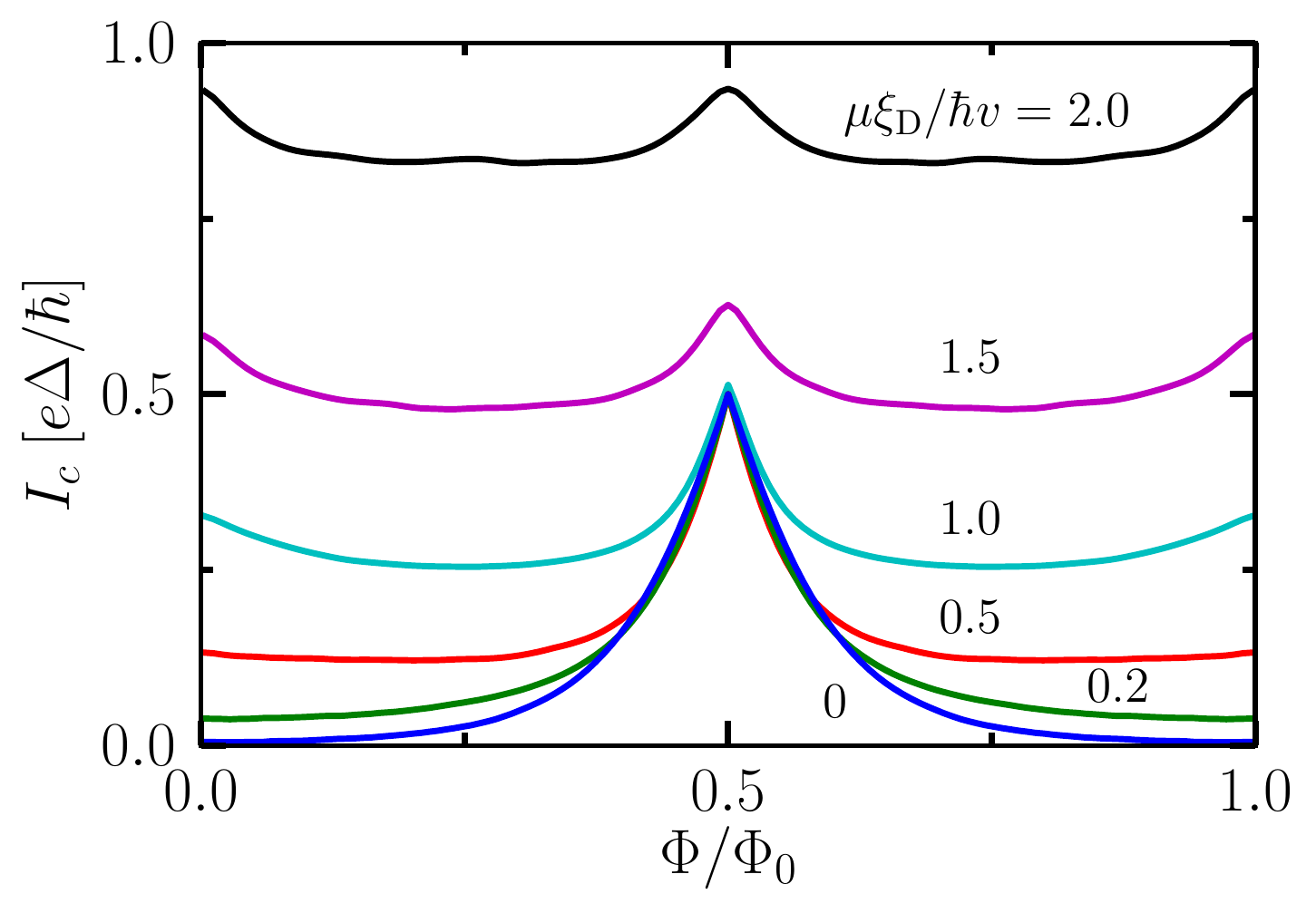}
\end{center}
\vspace{-0.5cm}
\caption{(Upper panel) critical current as a function of flux for $\mu\xi_\text{D}/\hbar v=0.5$ for various value of disorder. As disorder is increased, the peak sharpens. The inset shows the behavior at $\mu=0$, where disorder broadens the peak. (Lower panel) critical current as a function of flux for strong disorder $g=2$ and various values of chemical potential. Note the weak anti-localization peaks emerging at $\Phi=0,\Phi_0/2$ above $\mu\xi_\text{D}/\hbar v=1$.}
\label{fig:criticalDisordered}
\end{figure}

\begin{figure}[t!]
\begin{center}
\includegraphics[width = 0.95\columnwidth]{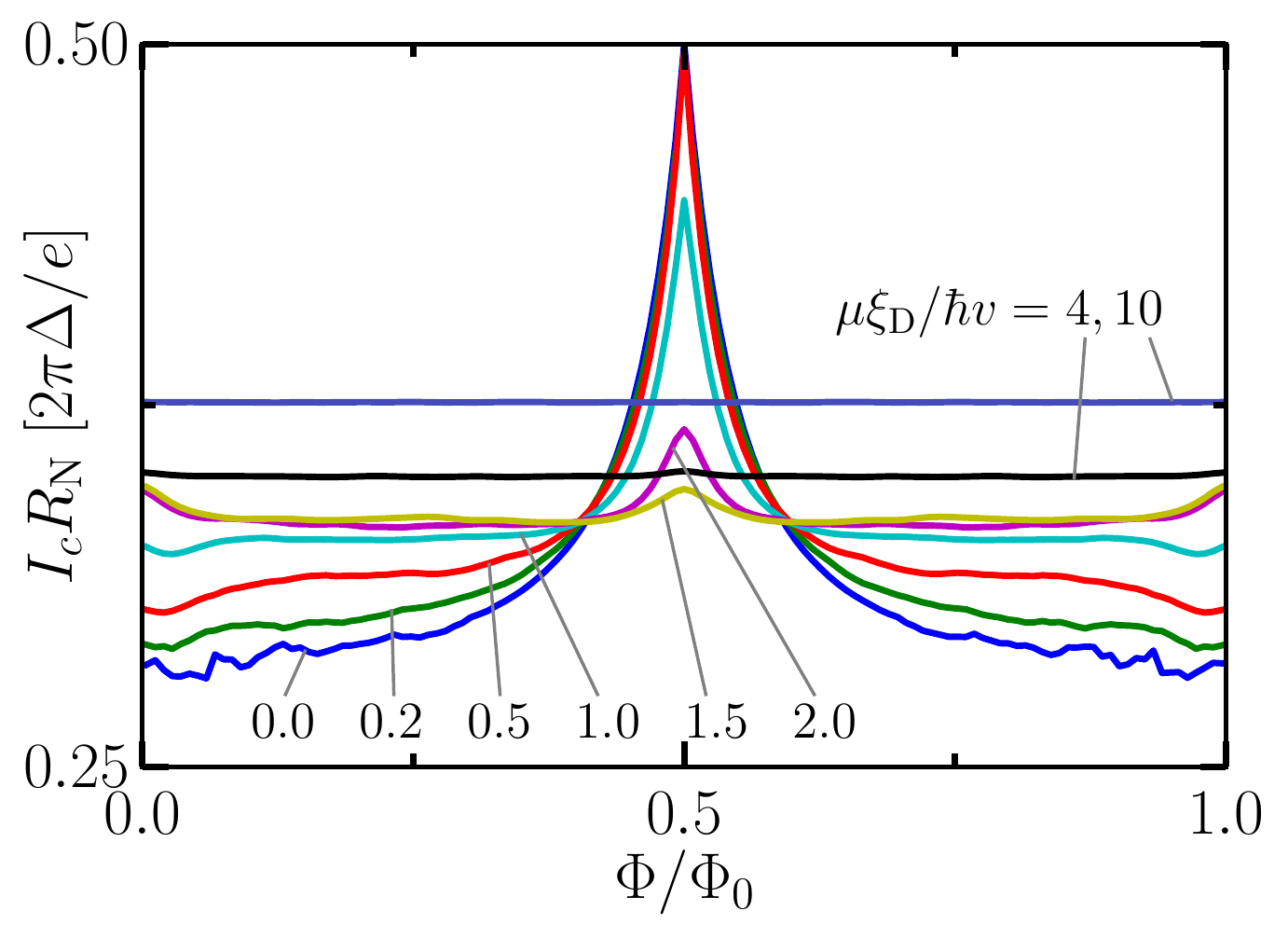}
\end{center}
\vspace{-0.5cm}
\caption{The product $I_CR_\text{N}$ as a function of flux for $g = 2$ and various values of chemical potential. At large chemical potential the product becomes independent of flux and saturates at a value close to $0.375$.}
\label{fig:IcRn}
\end{figure}

As another hallmark of the PTM appearing at $\Phi=\Phi_0/2$, note that one of the most striking differences between the  CPR at $\Phi=0$ and $\Phi=\Phi_0/2$ appears in the critical current at the Dirac point. At $\Phi=\Phi_0/2$ the critical current is finite, while at $\Phi=0$ it is exponentially small. It implies that the critical current as a function of $\Phi$ should peak at $\Phi=\Phi_0/2$. Fig.~\ref {fig:criticalDisordered} displays the behavior of the critical current as a function of flux for finite disorder strength and chemical potential. At low disorder, a broad distribution of current emerges, which then narrows and converges onto a sharp peak of height $e\Delta/2\hbar$ as the disorder strength is increased (note that at the Dirac point the behavior is slightly different: the sharp peak is broadened but still maintains an exponential form, see inset in upper panel of Fig.~\ref{fig:criticalDisordered}).  The lower panel of Fig.~\ref{fig:criticalDisordered} shows the critical current in the presence of strong disorder, $g=2$, for various values of $\mu\xi_\text{D}/\hbar v$~\footnote{For a comparison between weak and strong disorder see supplementary material.}. There is a wide range of chemical potentials for which the peak can still be clearly observed. Above $\mu\xi_\text{D}/\hbar v\sim1$, the critical current develops two peaks  at $\Phi=0$ and $\Phi_0/2$. These peaks are absent when plotting the product $I_cR_\text{N}$, as seen in Fig.~\ref{fig:IcRn}, where $R_\text{N}$ is the normal state resistance given by $R_\text{N}^{-1}=e^2/h\sum_n\tau_n$.

The appearance of a peak in the critical current and $I_cR_\text{N}$ at $\Phi=\Phi_0/2$ reflects the emergence of a PTM. The behavior at finite $\mu$ and $g$ can be understood as follows. According to Eq.~\eqref{eq:current}, for a clean wire at $\mu=0$, the critical current is expected to exhibit a sharp peak of height $e\Delta/2\hbar$ that drops exponentially as $\exp{(-L|\Phi/\Phi_0-1/2|/W})$. At finite $\mu$, the form of the current becomes slightly more involved: at first, the transmission probability of the lowest lying mode starts experiencing more resonances away from $\Phi=\Phi_0/2$. Then, other transmission probabilities become finite and start experiencing such resonances. Hence, for a finite $\mu$, $e\Delta/2\hbar$ represents the lowest bound for the peak height and its tails may no longer drop to zero. At low chemical potential, disorder localizes all modes apart from the PTM and suppresses the resonances in all modes away from $\Phi=\Phi_0/2$, recovering the sharp exponential peak.

At large values of the chemical potential, the two peaks in $I_c(\Phi)$ developing around $\Phi=0$ and $\Phi=\Phi_0/2$ are a direct result of time reversal symmetry recovered at the surface. This is manifested as an increase of the transmission probabilities $\tau_n$ due to quantum interference of time reversed paths, i.e, weak anti-localization.  Evidently, effects associated with weak anti-localization and the peak associated with the PTM mode can be disentangled from one another by considering $I_cR_\text{N}$ in which weak anti-localization features are cancelled out.  Hence, an appearance of a peak at $\Phi=\Phi_0/2$ in $I_cR_\text{N}$ is a genuine hallmark of the PTM.

The above calculations show that signatures of the PTM are found in the CPR, in the critical current as a function of flux, and in the product of $I_cR_\text{N}$ as a function of flux; we now comment on the effects of stray quasiparticles and finite temperatures.  In practice, fermion parity is unlikely to be conserved, and we expect that a supercurrent with a period of $2\pi$ will appear~\footnote{In fact, the $4\pi$ periodic CPR could also be compromised by the low energy branches in Eq.\eqref{eq:majorana branch} approaching $\Delta$ at $\phi=2\pi m$. This may result in a decay from the upper to the lower branch even if the crossings at $(2m+1)\pi$ are protected. We thank Gil Refael for this observation.}. Hence the discontinuity of the current at low chemical potentials for $\Phi=\Phi_0/2$ is the most promising unconventional effect in the CPR.  As the critical current is unaffected by fermion parity, such considerations are irrelevant for observing the peak at $\Phi_0/2$. At small but finite temperatures, transitions between the two energy branches Eq.~\eqref{eq:majorana branch} are expected to occur around $\phi=\pi$~\cite{FractionalJE}. Since the current coming from the two branches is the same in amplitude but opposite in sign, such transitions will average the current to zero near $\phi=\pi$. This will soften the discontinuity in the current and render the maximum of the current contributed by the PTM temperature-dependent, and consequently also the height of the peak in the critical current. Assuming the temperature is kept much smaller than $\Delta$, the maximal current coming from the PTM will roughly go from $e\Delta/2\hbar$ at $\phi=\pi$ to $e\Delta\sqrt{1-(T/\Delta)^2}/2\hbar$ slightly away from $\phi=\pi$ ~\footnote{A more precise estimate can be drawn from Ref.~\onlinecite{FractionalJE} using the full functional form of the critical current coming from the low energy branches at finite $T$. The actual power of $T/\Delta$ is slightly lower than $2$ at small $T/\Delta$}.   

These predictions should be observable in experiments with currently available materials. For the short junction limit to be met, the superconducting coherence length $\xi$ should be larger than the separation between the superconductors $L$. Estimates from experiments show that  $\xi$ can reach several hundreds of nanometers~\cite{PhysRevLett.109.056803}, which sets an upper bound for $L$. Ideally, the ratio $L/W$ should be kept smaller then unity, while $W$ should allow for reasonable flux values without destroying superconductivity. For a realistic value of  $W=400~nm$~\cite{Tian:2013kw,doi:10.1021/nl903663a}, threading a flux of $\Phi_0/2$ through the cylinder requires a field strength of about $0.2~T$. The disorder correlation length and the disorder strength are estimated to be $\xi_\text{D}\approx~10~nm$ and $g=0.5-1$~\cite{Beidenkopf}. Hence, the crossover value of $\mu\xi_\text{D}/\hbar v$ into the range where many modes acquire finite transmission corresponds to a density of $n\sim 8\times 10^{10}~cm^{-2}$, which is close to densities obtained in gated TI structures~\cite{Kim:2012eg}. Recent experiments have been carried out at temperatures $T \approx 30~m\text{K}$, while the gap size is estimated at about $\Delta=1~\text{K}$, rendering corrections to the peak height small. 

Finally, we discuss another motivation for studying 3DTI nanowires. Recently, JJs on the surface of 3DTI thin films were studied experimentally~\cite{PhysRevLett.109.056803,Sacepe,VenDerWiel,2012arXiv1209.5830C}. Typical length scales for devices made out of thin films are $L\sim 20-80~nm$, $W\sim0.5-3.2~\mu m$\cite{PhysRevLett.109.056803}. For such small values of $L/W$ we expect that effects associated with the PTM are overwhelmed by those contributed by other modes and difficult to resolve. In particular, we find that the Fraunhofer diffraction pattern of the critical current in the presence of a perpendicular field it is identical to the standard pattern~\footnote{See supplementary material.}. Hence such wide junctions may not be an ideal  arena to realize theoretical proposals targeting Majorana modes. In contrast, the nanowire geometry allows access to signatures of those low energy degrees of freedom while taking into consideration realistic conditions such as disorder, finite chemical potential, and temperature.

The authors would like to thank Kathryn Moler, Ilya Sochnikov, Katja Nowack, David Goldhaber-Gordon and Gil Refael for useful discussions. RI is an Awardee of the Weizmann Institute of Science -- National Postdoctoral Award Program for Advancing Women in Science. The authors acknowledges support from AFOSR (R.I.) and DARPA (J.H.B., H.-S.S., and J.E.M.). HSS is supported by Korea NRF (grant No. 2012S1A2A1A01030312).

\bibliography{JJ2}
\clearpage
\appendix*
\section{Supplementary Material}

\section{Comparing the effect of weak and strong disorder on the critical current as a function of flux through the wire}

It is shown in the main text that disorder in 3D topological insulator (3DTI) nanowires has an effect on the critical current through a Josephson junction (JJ) as a function of flux.  Due to the fragile nature of modes unprotected by time reversal symmetry, disorder leads to localization which sharpens the features of current that result from the perfectly transmitted mode (PTM). Hence we expect that for stronger disorder, it would be possible to observe the exponential peak more clearly up to higher chemical potentials.  This section compares the effects of weak vs. strong disorder to validate our prediction.

\begin{figure}[t!]
\includegraphics[width=1\columnwidth]{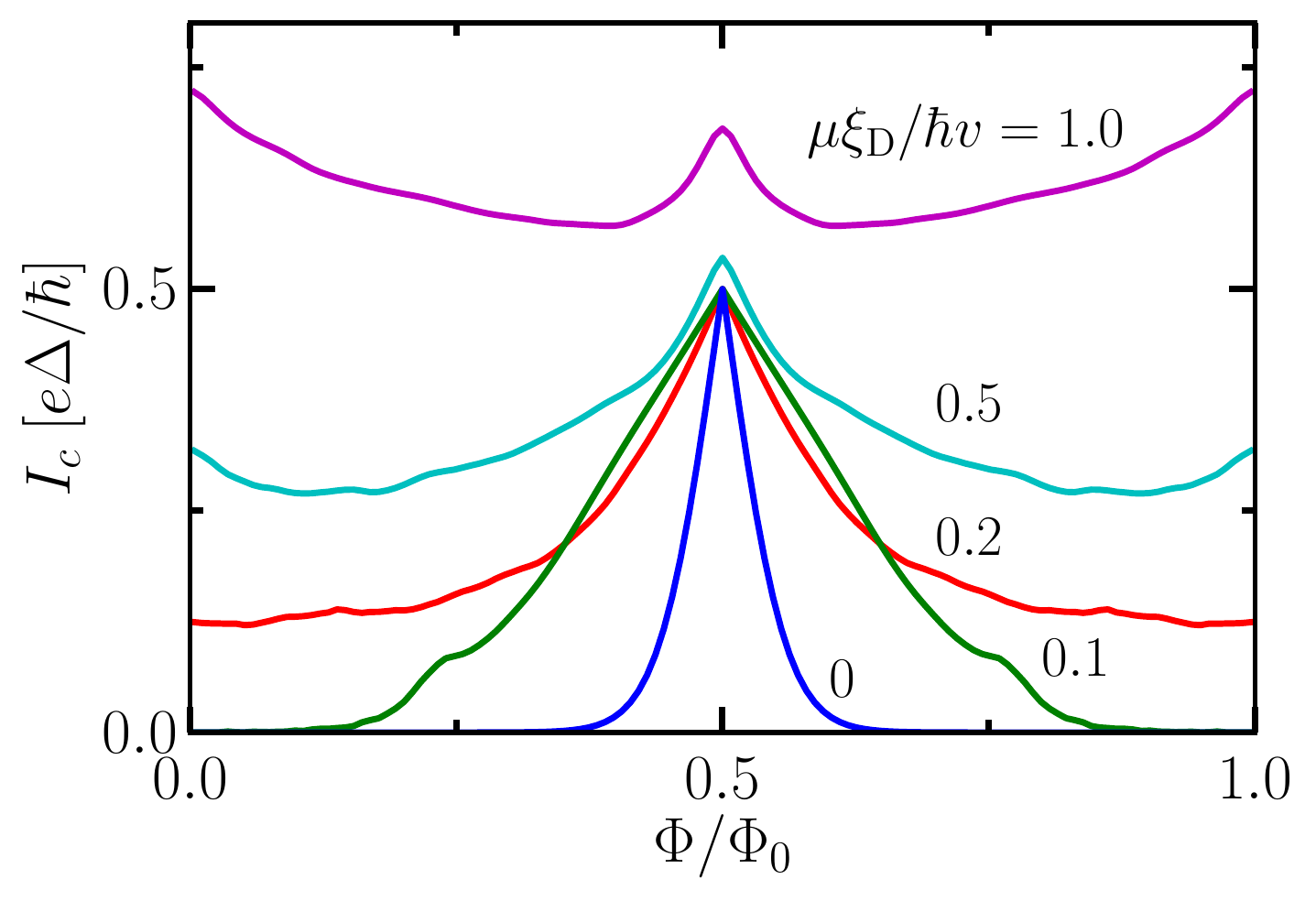}
\includegraphics[width=1\columnwidth]{peak_g2p0.pdf}
\caption{ Critical current as a function of flux for weak disorder $g=0.5$ (upper panel) vs. strong disorder $g=2$ (lower panel). Note the weak anti-localization peaks emerging at $\Phi=0,\Phi_0/2$ above $\mu\xi/\hbar v=1$ for strong disorder, while for week disorder those emerge already above $\mu\xi/\hbar v= 0.2 $.}
\label{fig:disorder}
\end{figure}

Fig.~\ref{fig:disorder} contrasts the critical current for $g=0.5$ and $g=2$ (the lower panel is identical to one appearing in Fig.~(3) of the main text). As it shows, qualitatively the same effects emerge for both disorder strengths, although the sharp peak survives to lower values of the chemical potentials with stronger disorder. For $g=0.5$, the exponential peak disappears at about $\mu \xi_\text{D}/\hbar v\sim 0.2$, while for $g=2$ it survives up to $\mu \xi_\text{D}/\hbar v\sim 1$.

\section{Fraunhofer diffraction pattern for wide Josephson Junctions on 3D topological insulator films}

One of the most common measurements done on planar Josephson junctions is that of the critical current as a function of flux through the junction. The critical current displays a diffraction pattern, which for standard junctions with uniform current injection between two S-wave superconductors takes the ``Fraunhofer'' form $I_c(\Phi)/I_c(0)=|\sin{(\pi\Phi/\Phi_0)}/(\pi\Phi/\Phi_0)|$~\cite{Tinkham}.

  \begin{figure}[t!]
\includegraphics[width=1\columnwidth]{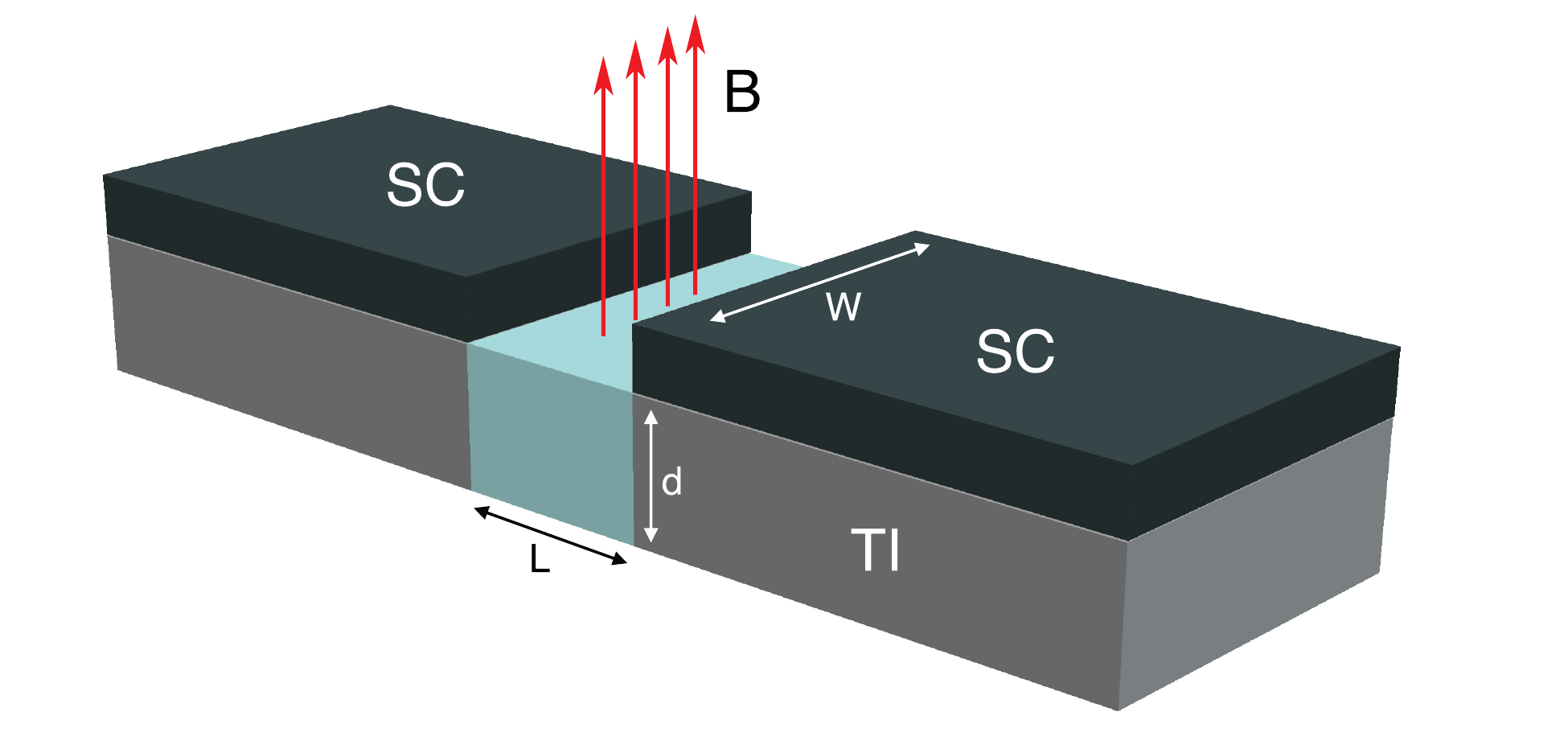}
\includegraphics[width=1\columnwidth]{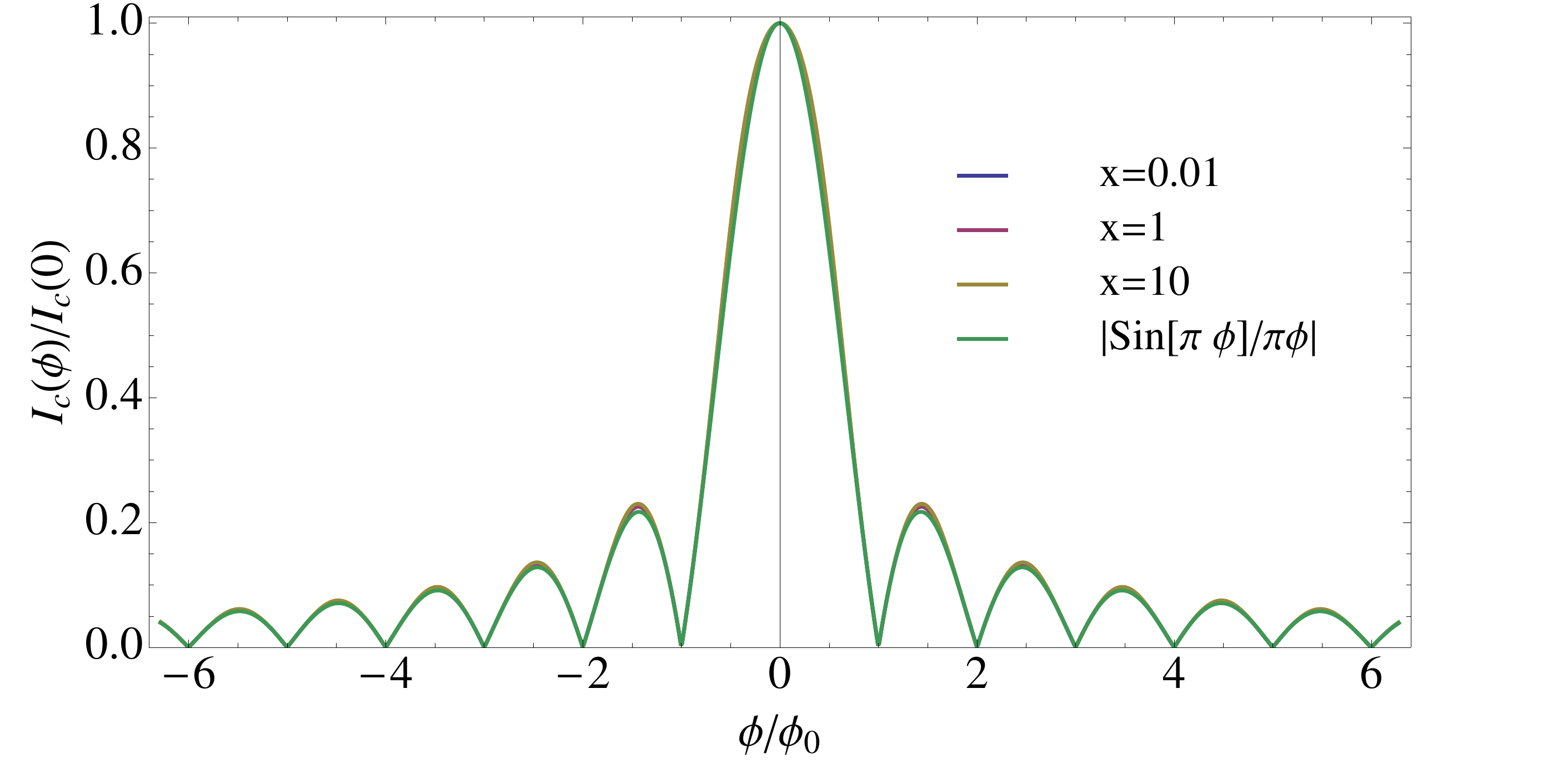}
\caption{Top: Josehson junctions on the surface of 3D TI thin film. Bottom: Fraunhofer diffraction pattern for a wide Josephson junction on the surface of a TI film. Here, $L/W=0.1$ and $x=\mu L/\hbar v$. The curves are rescaled by $I_c(0)$, exhibiting a perfect collapse. }
\label{fig:setup2}
\end{figure}

To consider the effect of a magnetic flux penetrating a wide JJ junction situated on a thin film of a 3DTI, we generalize Eqn.(2) and (3) of the main text to include the effect of the flux via a semi-classical approach. The effect of the field on the modes in the normal part is neglected such that the current and the transmission probabilities remain the same, while the phase difference between the superconductor gets shifted and becomes $y$ dependent, $\phi\rightarrow \phi-2\pi\frac{\Phi}{\phi_0}\frac{y}{W}$, with $\Phi=BLW$ is the total flux through the junction. Such an approximation is valid provided the length of the junction is much smaller then the magnetic length $\ell_B=\sqrt{e\hbar/B}$. For current experimental systems and conditions $\ell_B>0.5\mu$, while $L$ is of the order of tens of nanometers, hence the semiclassical approach is appropriate and should capture the physics those junctions display.  The $y$ dependence of the superconducting order parameter translates into a $y$ dependent Josephson current, and the total current is obtained by integrating over that current across the junction. 

For the case of a wide junction ($W>>L$) the calculation becomes independent of the choice of boundary conditions determining the values of the momenta $q_n$. We would like to pause here and note the implications of that: in the nanowire geometry extensively discussed in the main text, the boundary conditions had a crucial role in transport, and their manipulation determined whether or not a PTM will be present. Here  they play no role, explaining the insignificant role the PTM  plays in transport within wide junctions. Here we choose $q_n=\pi/W(n+1/2), n=0,1,2...$ corresponding to the so called infinite mass boundary condition\cite{PhysRevB.74.041401}.  The Fraunhofer interference pattern as a function of flux is presented in Fig.\ref{fig:setup2}. On top of being identical to the standard pattern, it also appears not to depend on the chemical potential (patterns corresponding to deferent values of the the chemical potential collapse when rescaled by $I_c(\Phi=0)$. 

Fraunhofer patterns in JJ on 3DTIs were measured by several groups~\cite{2012arXiv1209.5830C,VenDerWiel,PhysRevLett.109.056803}. Note that in some measurements already performed in 3DTI, deviations from the standard pattern were observed~\cite{VenDerWiel,PhysRevLett.109.056803}. For example, for some measurements the minima of the interference pattern do not approach zero~\cite{PhysRevLett.109.056803}, but in fact a finite current is measured.  A possible explanation is that these residual currents might be flowing through the side surfaces of the thin film (see Fig.~\ref{fig:setup2}) or result from some other inhomogeneous current pattern. For these side surfaces, the shift in the phase of the order parameter is constant (since the field is parallel to the surface), and hence these will not display an interference pattern, but rather contribute an amount of current which is periodic in $\phi$. Since that piece of the current will not show a diffraction pattern, it will contribute to lifting the nodes of the pattern coming from the top and bottom surfaces. This conjecture can be checked in experiment by tilting the field, thereby having the flux penetrating the side surfaces as well. Other deviations from the standard pattern include a periodicity that is different than a single flux quantum, as well as irregular spacings of the zeros of the diffraction pattern, were reported~\cite{VenDerWiel,PhysRevLett.109.056803}. While we find a pattern for which the minima appear at integer multiples of $\phi_0$, it is possible that different length scales in the system (such as the penetration depth for example) are field dependent. Such physics is beyond the scope of our calculations.

\end{document}